# DFQ--Double Frequency RFQ

Y. Iwashita

NSRF, ICR, Kyoto University, Gokanosho, Uji, Kyoto 611-0011 JAPAN

*Abstract*

RFQ with a harmonic higher order quadrupole mode is studied. Assuming that we superpose a higher order mode with twice the frequency of the fundamental mode, a sawtooth waveform is approximated. In such a case, the bunching function is enhanced while the transverse stability is modified. Second order longitudinal harmonic component is also required to enable the effect.

## 1 INTRODUCTION

Some bunchers use sawtooth wave forms for better bunching factor. Application of such sawtooth wave form to an RFQ may shorten a length of RFQ. Because of the orthogonality of the trigonometric functions, we also have to introduce a longitudinal harmonics. Energy gain in a period is described in the next section. Section 3 shows a potential function generating such electric field and an example of its vane shape. RF defocusing factor is calculated in section 4. The transverse stability is discussed in section 5. Rough resonator examples can be seen in section 6.

## 2 BUNCHING ENHANCEMENT

Let us assume that the electric field $E_z$ on a beam axis in an RFQ cell with length of $L_c$ is expressed as:

$$E_z(t,z) = E_0\left(\sin(\omega t + \phi) + H_i \sin(i(\omega t + \phi + \phi_i))\right) \left(\sin(kz) + jA_{j0}\sin(j(kz + z_j))\right), \quad (1)$$

where $\omega$ is angular frequency of RF and $k = \pi/L_c$. The second terms of time and spatial harmonics are newly introduced in addition to the two term potential function [1,2]. The energy gain $\delta T$ for an ion with relationship $\omega t = kz$ is obtained by integrating Eq.(1) from $z=0$ to $2L_c$:

$$\delta T = q \int_0^{2L_c} E_z(kz/\omega, z)\,dz, \quad (2)$$

where $q$ is the charge of the ion. Because of the orthogonality, we need a condition $i=j$ for the harmonics effect. In order to enhance the bunching function in an RFQ, a condition $i=j=2$ will be taken. It should be noted that the range of the integration is twice the range of the conventional RFQ case. Thus, the energy gain in the period is:

$$\delta T_2 = qE_0\left(\cos\phi + 2A_{20}H_2\cos(2(\phi + \phi_2 - \pi z_2))\right). \quad (3)$$

In order to approximate the sawtooth function, $\phi_2 - \pi z_2$ and $A_{20}H_2$ have to be $-\pi/4$ and $1/4$, respectively. Figure 1 shows the resulted energy gain as a function of $\phi$.

## 3 POTENTIAL FUNCTION

The potential function that generates such longitudinal harmonics is expressed by:

$$U(r,\psi,z) = \frac{V}{2}\left\{\left(\frac{r}{r_0}\right)^2 \cos 2\psi + A\left(I_0(kr)\cos(kz) + A_{20}I_0(2kr)\cos(2kz + z_2)\right)\right\} \quad (4)$$

where $A = \dfrac{m^2 - 1}{m^2 I_0(ka) + I_0(mka)}$ and $a$ is the minimum radius at $z=0$[1]. The electric fields are:

$$E_r = -\frac{V}{r_0^2} r \cos 2\psi - \frac{kAV}{2}\{I_1(kr)\cos kz - 2A_{20}I_1(2kr)\cos(2k(z+z_2))\},$$

$$E_\psi = \frac{V}{r_0^2} r \sin 2\psi,$$

$$E_z = \frac{kAV}{2}\left(I_0(kr)\sin kz + 2A_{20}I_0(2kr)\sin(2k(z+z_2))\right). \quad (5)$$

Cross sections of vanes can be obtained by contours:

$$U(x,0,z) = \frac{V}{2}\text{ for x vane and }U\left(y,\frac{\pi}{2},z\right) = \frac{V}{2}\text{ for y vane.}$$

An example of the vane shapes with $L_c/r_0=1.5$, $A=0.05$, $A_{20}=0.25$ and $z_n=0$ are shown in Fig. 2. Because of the second order term, x and y vanes are not symmetric. It can be seen that y-vane is separated at z=0 and z=3$L_c$, which should not be harmful in this particular case, because of the small distance. The condition $A_{20}H_2=1/4$ implies $H_2=1$. Figure 3a shows the time dependent term of Eq.(1). Figure 3b shows the corresponding longitudinal distribution. It should be noted that the peak value is higher than the single sine wave, which will be discussed later. Larger $A_{20}$ term in short $L_c$ region makes the contour lines complex, which means that the $A_{20}$ cannot be large at such region. As shown in Fig.2, the vane tip position $sa$ at (y,z)=(0,0), is modified from the original value $a$ by the $A_{20}$ term. $A_{20}$ is plotted as functions of the coefficients at $m$=1.1, $L_c/r_0$=1.1,1,2, and 1.5 in Fig. 4, where $m$ is the commonly used modulation parameter.

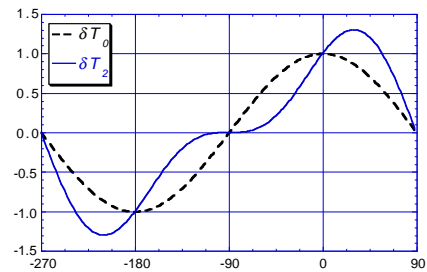

Fig.1 Approximated sawtooth energy gain function $\delta T_2$ and the original function $\delta T_0$.

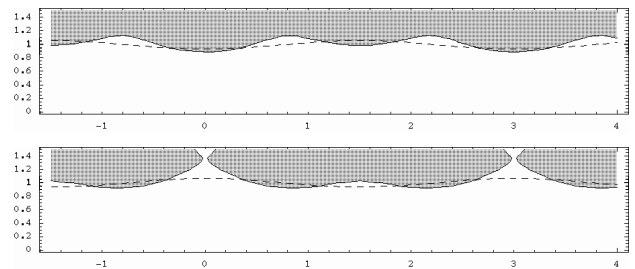

Fig. 2 Cross sections of x- and y-vanes. Horizontal and vertical axes denote $z$ and $r/r_0$, respectively. $L_c/r_0=1.5$, $z_2=0$, $A=.05$, $A_{20}=.25$. Broken line shows the ordinary vanes.

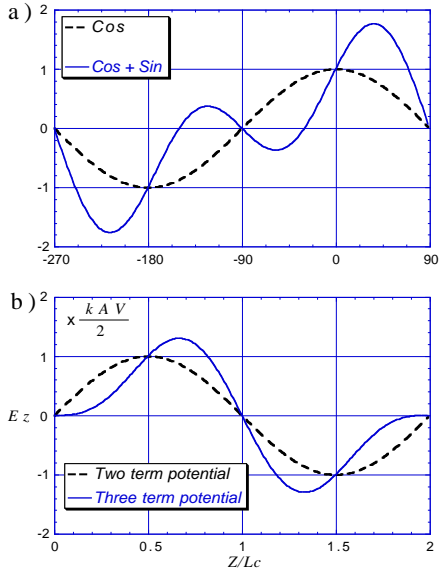

Fig. 3  a) Time dependent term in the example, b) the longitudinal distribution.

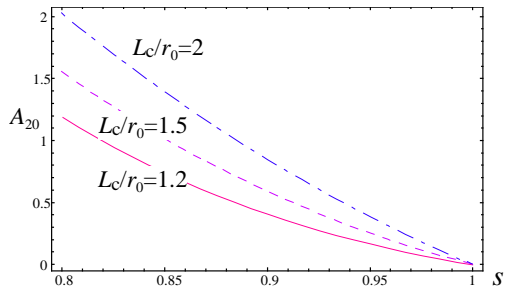

Fig. 4   $A_{20}$ as functions of $s$.

## 4  RF  DEFOCUSING

The averaged RF defocusing force $D_x$ is obtained by integrating $E_x$ with time dependent factor over a period. $E_x$ and $E_y$ are given by:

$$\begin{pmatrix}E_x\\E_y\end{pmatrix}=\begin{pmatrix}\cos\psi & -\sin\psi\\\sin\psi & \cos\psi\end{pmatrix}\begin{pmatrix}E_r\\E_\psi\end{pmatrix}. \quad (6)$$

Thus, $D_x$ is given by:

$$D_x=\frac{q\int_0^{2L_c}E_x\big(\sin(kz+\phi)+H_2\sin(2(kz+\phi+\phi_2))\big)dz}{2L_c}$$
$$=-\frac{qk^2AVx}{4}\big(\sin\phi-4A_{20}H_2\sin(2kz_2-2\phi-2\phi_2)\big), \quad (7)$$

where the $E_x$ function are approximated up to first order of x around the beam axis. $D_y$ is given by the same form as $D_x$. Figure 5 shows $D_x$ as a function of $\phi$. The RF defocusing term is zero at the synchronous phase of 90° (bunching operation) because of the flat region in $\delta T_2$.

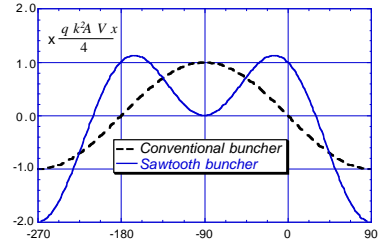

Fig. 5   RF defocusing terms for the approximated sawtooth function and the conventional sine function.

## 5  TRANSVERSE  STABILITY

The stability of the transverse motion is obtained by Hill's equation:

$$\frac{d^2x}{d\eta^2}+\big(B(\cos 2\pi\eta+A_{20}\sin 4\pi\eta)+\Delta\big)x=0,$$

where  $\eta=\frac{z}{2\beta\lambda}$,  $B=\frac{q\lambda^2 V}{m_0c^2r_0^2}$. (8)

Figure 6 shows the stability regions for cases of $A_{20}$=1, 1/2, 0 and $A_{20}$=1 with B scaled by 71%. The last one corresponds to preserving the total RF power with assumption of the equal shunt impedance for the second order mode. Because of the extra force term, the focusing strength increases with $A_{20}$ for constant amplitude of the fundamental mode.

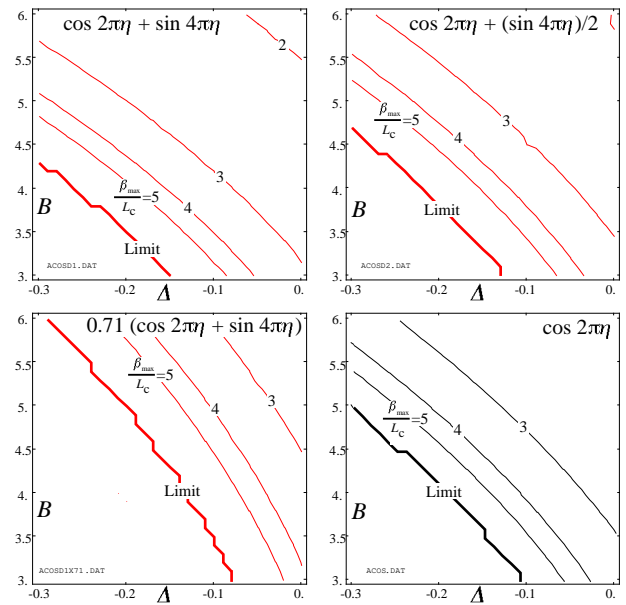

Fig. 6   Stability diagram for $A_{20}$=1,1/,1/3 and 0.

## 6 CAVITY EXAMPLE

Figure 7 shows an example of the harmonic cavity for a four-vane DFQ. The right lower half of the figure shows the fundamental mode and the left upper one shows the second order mode. This particular example exhibits rather small shunt impedance (less than half) compared with that of a conventional single-mode-RFQ[3]. Figure 8 shows another example for four-rod-DFQ that has the frequencies of 143 and 290MHz. These are shown only for the possibility of the harmonic resonators, where the geometries are not fully optimized.

The equivalent circuit for such resonators is shown in Fig. 9. Using following notations:

$$\omega_1 = \frac{1}{\sqrt{L_1 C_{gap}}}, \omega_2 = \frac{1}{\sqrt{L_2 C_2}}, u = \frac{L_2}{L_1} \text{ and } w = \frac{\omega_2}{\omega_1}, \quad (9)$$

and condition that the ratio of two resonant frequencies is two, $u$ and $w$ should hold following equation:

$$w = \left(5 + \sqrt{9 - 16u}\right) / \left(4(1+u)\right). \quad (10)$$

The current ratios $i_1/i_2$ for both the modes is shown in Fig.10. $u$ should be chosen with power supply specifications. This knowledge would be helpful for the design of a real cavity.

One more scheme is to apply fundamental RF to horizontal electrodes and to apply second order mode to another ones. This scheme is under investigation.

## 7 DISCUSSION

$L_C/r_0$ at an entrance is large for a low frequency RFQ (as used for heavy ions), which allows larger $A_{20}$ value. The sparking issue in the superposed RF wave form is not clear, but seems easier for lower frequencies [4,5,6].

The complex vane shapes may be approximated by trapezoids, because only the longitudinal higher order mode having the corresponding RF mode can affect the longitudinal motion. The transverse motion will not be changed much as long as the quadrupole component is preserved. This can be extended to an IH-DTL (inter-digital H type) with electro-focusing fingers, where the gap centers shift alternatively.

Because of the wide stability region in the synchrotron oscillation, the synchronous phase may be at 0° or more, which makes the accelerator section short. The RF power consumption is just a sum of that for each mode, in spite of the high peak field. Because two additional parameters are added (phase and amplitude for the second RF), focusing characteristics may be adjusted independently. If z-dependence in the quadrupole field is added, more focusing force may be available with penalty of multipole effects. The determination of the cell parameters is complex compared with the conventional RFQ. The multiple RF feed technique is to be established[7].

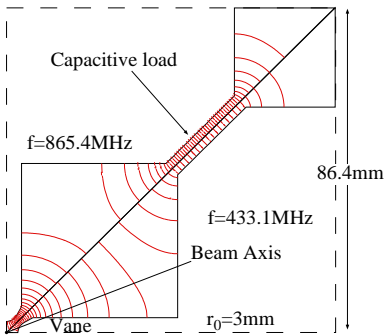

Fig. 7 An example of harmonic cavity for a four-vane DFQ. Beam axis is bottom left.

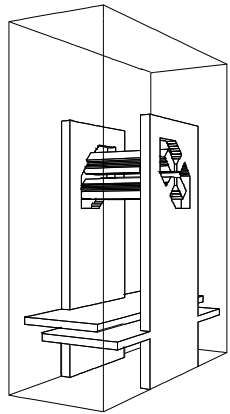

Fig. 8 4-rod-DFQ

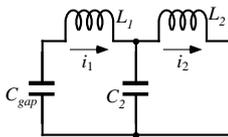

Fig. 9 The equivalent circuit for the DFQ

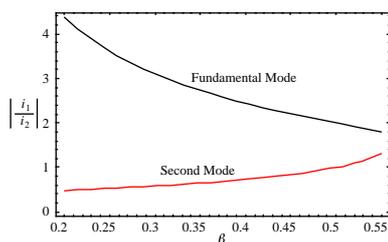

Fig. 10 Current ratios $i_1/i_2$ as functions of $u$.